\documentclass[10pt, a4paper]{article}
\usepackage[utf8]{inputenc}
\usepackage{authblk}
\usepackage{booktabs}
\usepackage{cite}
\usepackage{fullpage}
\usepackage{amsmath}
\usepackage{graphicx}
\usepackage{tabularx}
\usepackage{url}
\usepackage{titlesec}
\usepackage{multirow}

\title{Detecting absence: A dedicated prediction-error signal emerging in the auditory thalamus}
\author[1,2,3,5*]{Alejandro Tabas}
\author[4,5]{Heike S\"{o}nnichsen}
\author[5]{Sandeep Kaur}
\author[6]{Marco Meixner}
\author[3,5]{Katharina von Kriegstein}
\affil[1]{Basque Center on Cognition, Brain and Language; San Sebastian, Spain}
\affil[2]{Ikerbasque, Basque Foundation for Science; Bilbao, Spain}
\affil[3]{Max Planck Institute for Human Cognitive and Brain Sciences, Leipzig, Germany}
\affil[4]{Institut f\"ur Psychologie I, Universit\"at zu L\"ubeck, L\"ubeck, Germany}
\affil[5]{Chair of Cognitive and Clinical Neuroscience, Technische Universit\"{a}t Dresden, Dresden, Germany}
\affil[6]{Ruhr-Universit\"at Bochum, Bochum, Germany}
\affil[ ]{\normalfont *Corresponding author: \protect\url{tabas@bcbl.eu}}
\date{}
\titleformat*{\section}{\large\bfseries}
\titleformat*{\subsection}{\bfseries}

\begin{document}

  \maketitle

  \section*{Abstract}
    How does the brain know what is out there and what is not? Living organisms cannot rely solely on sensory signals for perception because they are noisy and ambiguous. To form stable percepts, the brain exploits its prior knowledge or beliefs. Current theories explain how perceptual beliefs on the features of the stimuli (e.g., pitch or motion velocity) are updated by feature prediction errors: the mismatch between expected and observed values. How the brain updates beliefs about a stimulus' presence or absence is, however, unclear.

Previous work has argued that detecting absence, which relies on the lack of evidence for presence, depends on different mechanisms than the identification of a feature. Here, we tested the hypothesis that the detection of absence relies on a form of prediction error dedicated to the update of beliefs on stimulus occurrence. Since feature prediction error already occurs in the sensory pathway, we hypothesised that this would be similar for absence prediction error. We studied this absence prediction error in the human auditory pathway with fMRI by measuring BOLD responses to absence of stimuli for which the participants held different levels of expectations.

Comparing more and less expected responses to absence, we showed that prediction error to absence is represented in the auditory thalamus and cortex. Moreover, while feature prediction error is encoded in the auditory midbrain, absence prediction error was not, implying that absence is processed by a different circuit. These results identify the neural mechanism for the detection of sensory absence. These mechanisms may be disrupted in conditions like psychosis, where the detection of absence is impaired.

  \section*{Introduction}
    
To survive, living organisms need to identify not only what is present but also what is absent. They do so by relying on noisy sensory inputs that often lack an unambiguous interpretation. Consequently, organisms draw on their previous knowledge of the sensory world to interpret sensory information \cite{Aitchison2017}. While such knowledge facilitates encoding and helps process expected inputs \cite{Chandrasekaran2009, Kok2012, Font2020}, it can also lead to incorrect perception of elements that are not actually there \cite{Ford2012, Beck2014, Sterzer2018}. Understanding how knowledge and sensory input interact in the identification of perceptual objects is essential for explaining normal perceptual function and its failures, such as illusions and hallucinations.

Previous work has extensively investigated how the brain identifies the correct feature of a stimulus (e.g., the pitch of a tone) \cite{Rao1999, Friston2003, Friston2009, Aitchison2017, Keller2018, DeLange2018, Tabas2021}. Current theories of perception propose that relevant prior knowledge is compressed into a prediction: a probabilistic representation of the prior beliefs represented by the mean and variance of the feature \cite{Friston2003, Friston2009, Millidge2022, Hertag2025}. Sensory systems compare this prediction with incoming sensory inputs to compute prediction error: the difference between the two means, weighted by their relative variances. Prediction errors then update the prior beliefs into a posterior representation of the perceived feature. These errors can be positive, if they increase the magnitude of the feature (e.g., shifting to a higher pitch value) or negative, if they decrease it \cite{Keller2018, Hertag2025}. The encoding of both positive and negative \cite{Jordan2020, Hertag2025, Leonardon2025} prediction errors has been thoroughly demonstrated across perceptual systems and species (see~\cite{Heilbron2017, Walsh2020, Antunes2021} for reviews). However, it remains unclear how prediction errors are used to update beliefs when the expected stimulus does not occur at all.

Here we propose that there is a separate kind of probabilistic representation that encodes beliefs about the presence or absence of a stimulus, which is updated by a dedicated kind of prediction error. We hypothesise that reductions on presence plausibility might be mediated by what we call absence prediction error. Previous work did not consider absence prediction error \cite{Heilbron2017, Walsh2020, Tabas2021} and confounded feature prediction error with absence prediction error. For instance, oddball paradigms, the gold standard in the study of prediction error, induce predictions by regularly presenting a standard (e.g., a tone), which is rarely interrupted by a deviant (e.g., a tone of a different pitch). Prediction error to deviants can be understood as feature (shifting the mean of the pitch from the standard to the deviant value), or absence (increasing the plausibility that the deviant pitch is present while decreasing the plausibility that the standard pitch is present) prediction error. These paradigms have therefore intrinsically confounded both error types. 

Prediction error to unexpected absence has been investigated mainly through the so-called \emph{omission paradigm} (see~\cite{Braga2022} for a review), where a periodically presented standard stimulus is occasionally replaced by silence (i.e., an omission). The \emph{omission responses} evoked in this paradigm are often interpreted as prediction errors \cite{Braga2022, Wacongne2011, Bendixen2012}. However, omission responses may also result from neural entrainment to the periodic presentation of stimuli \cite{Will2007, Heilbron2017, LHermite2023}. 

In this preregistered study \cite{Tabas2022prereg} we used a novel paradigm that avoids potential neural entrainment confounds by introducing different levels of expectations. These expectations were initiated by abstract rules that modulated the prior beliefs of the participants on the presence/absence of auditory stimuli. We measured responses to stimulus absence under different levels of expectation to determine whether the sensory pathway encodes absence prediction error (Hypothesis 1), and whether absence prediction error is represented within the same circuits where feature prediction error was previously reported. We chose the auditory pathway as a model system because there is strong evidence that feature prediction error is expressed along the ascending hierarchy \cite{Carbajal2018, Tabas2021, Parras2017}. In addition, we examined whether omission responses, previously reported only in the cerebral cortex \cite{Braga2022, Lao2023}, are also encoded in subcortical sensory pathways (Hypothesis 2). Understanding how sensory systems infer the presence/absence of stimuli is essential not only to comprehend general perceptual function, but also to devise new hypotheses for clinical conditions such as psychosis, where there may be specific difficulties in recognising the absence of predicted stimuli \cite{Sterzer2018, Keller2024}.
  
  \section*{Results}

\subsection*{An experimental paradigm to elicit absence prediction error}

\begin{figure}[tbh!]
    \centering
    \includegraphics[width=\textwidth]{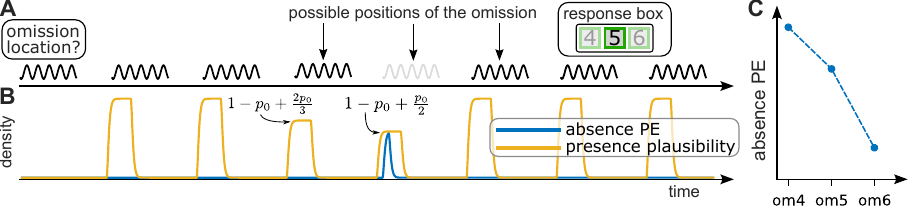}
    \caption{\textbf{Experimental design and hypotheses}. 
    A) Example of a trial. Each trial consisted of a sequence of up to eight repetitions of one pure tone (standards; black); in $p_0 = 6/7$ of the trials one of the pure tones was omitted (omission; light grey). In each trial, the omission could be located in positions 4, 5, or 6 of the sequence. Participants were instructed to report, via a button press, the position of the omission within the sequence. Tones and omissions within each trial were separated by 500\,ms inter-stimulus-intervals ($ISI = 500$\,ms). 
    B) Expected responses from regions encoding absence prediction error (aPE). We assume that the predicted plausibility is the optimal estimate of the probability of presence right after hearing the tone. aPE is defined as the prediction probability of the stimulus presence during an omission.
    C) If there are brain areas with absence PE, the response to the omissions should scale with the plausibility of the presence of the omitted tone. The plot shows the expected responses to the regressors corresponding to omissions in positions 4 (probability of presence of a tone $5/7$), 5 (probability $4/7$) and 6 (probability $1/7$) for regions encoding aPE.
    \label{fig:design}}
  \end{figure}

  We developed a new paradigm to identify brain areas encoding absence prediction error (aPE; Figure~\ref{fig:design}). To avoid confounding aPE with passive effects \cite{May2010} such as entrainment \cite{Will2007, LHermite2023}, we used abstract rules that elicit different levels of expectations on the omissions: any passive effects should be apparent for any omission, whereas aPE should be inversely proportional to the expectations on the presence of the omission. We then used 3-Tesla fMRI with a resolution of 1.5\,mm isotropic to measure responses in the inferior colliculus (IC, auditory midbrain), the medial geniculate body (MGB, auditory thalamus), and the primary auditory cortex (PAC) while participants engaged in the paradigm.
  
  A trial consisted of a sequence of up to eight repetitions of a pure tone separated by a constant inter-stimulus-interval (Figure~\ref{fig:design}A). Trials contained one omission (a missing pure tone) with probability $p_0 = 6/7$. We asked the participants to attend and report, via a button press, the position of the omission within the sequence, if any. Expectations on the omissions were modulated by three abstract rules (similar as in~\cite{Tabas2020}) that were disclosed to the participants: 1) trials can have only up to one omission; 2) some of the trials have no omissions; and 3) omissions can only occur in positions 4, 5, or 6. The three omission positions were used the same number of times throughout each experimental session, ensuring that the participants had similar expectations for each position at the beginning of each trial. This made the probability of finding an omission in position 4 after hearing 3 tones $p_0/3$. However, if the omission is not in position 4, it can only be placed in positions 5 or 6; this made the probability of finding an omission in position 5 after hearing 4 tones $p_0/2$. The probability of finding an omission in position 6 ($om6$) after hearing 5 tones is $p_0$. We assume that the beliefs of the participants on the plausibility of a tone is that of an ideal observer; namely, the probability that the sequence does not have an omission $(1-p_0)$ plus the probability that the sequence has an omission $p_0$ multiplied by the probability that the omission does not occur at that position $(1 - p_{\text{omission}})$. E.g., for position 4, $p_{\text{presence}} = (1-p_0) + p_0 * (1-1/3) =  1/7 + 6/7 \, (1 - 1/3) = 5/7$.
  
  We modelled the brain responses to the paradigm using six regressors (Supplementary Figure~S1C): responses to the first tone of a trial ($std0$); responses to the subsequent tones preceding either the omission or, in trials without omission, the sixth tone ($std1$); responses to tones following the omission or the sixth tone ($std2$); and responses to the omissions in positions 4, 5, or 6 ($om4$, $om5$, $om6$, respectively). Regions encoding aPE would selectively respond to omissions with an amplitude related to the expectations on the presence of the omitted tone (Figure~\ref{fig:design}C, red).

\subsection*{Responses to unexpected absence occur at all stages of the auditory pathway}

  First, we verified that auditory ROIs adapted to the repetition of the standard, as reported in previous studies \cite{Cacciaglia2015, Tabas2020, Tabas2024, Ara2024}. As expected, adaptation was robustly present in all ROIs (Figure~\ref{fig:omres}A); $p < 0.025, d \geq 0.98$; all effect sizes and $p$-values are listed in Supplementary Table~S1; all $p$ values listed in the results section are corrected for multiple comparisons - see Figure legends for more information).
  
  We then examined whether, as predicted by our second hypothesis, omission responses occurred in subcortical pathways. In agreement with the hypothesis, we observed stronger BOLD responses to unexpected omission $om4$ than to repeated tones $std1$ in all auditory ROIs (Figure~\ref{fig:omres}B). Responses to the omissions were significant in bilateral PAC (Figure~\ref{fig:omres}B; $p \leq 10^{-4}, d \geq 2.2$), MGB ($p < 10^{-3}, d \geq 1.3)$), but not in IC ($d \leq 0.74$). On visual inspection, the responses increased from IC to MGB to PAC.

  \begin{figure}[bth!]
    \centering
    \includegraphics[width=\textwidth]{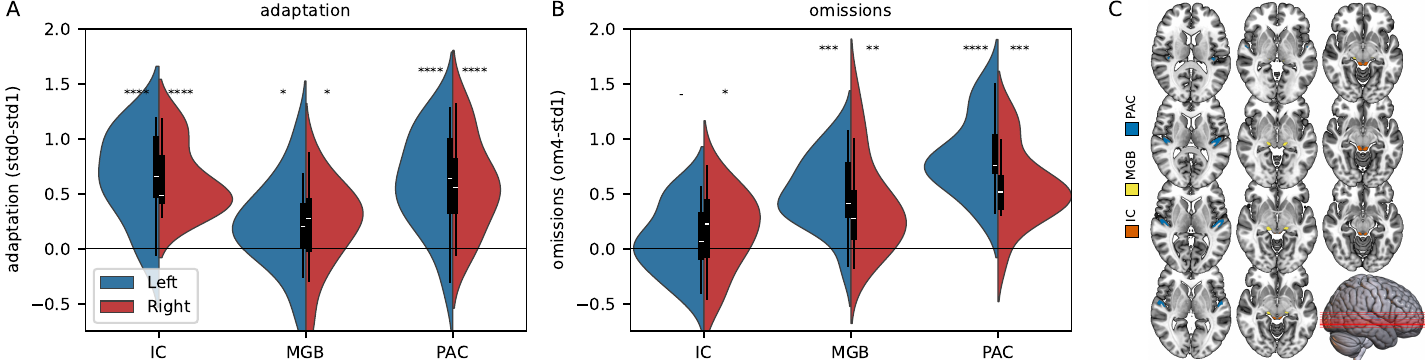}
    \caption{\textbf{Omission responses and adaptation}. A) Adaptation strength, as measured by the contrast $std0 - std1$, where $std0$ is the first tone of a trial and $std1$ corresponds to the tones before the omission or the sixth position with no omission. B) Strength of the response to omissions as measured by the contrast $om4 - std1$. Blue halves of the violin plots correspond to the left hemisphere; red halves to the right hemisphere. Violin plots are kernel density estimations across the entire population (one sample per participant). All $p$-values were computed using one-sided ranksum tests and Holm-Bonferroni corrected for 12 comparisons (nominal corrected and uncorrected $p$-values are listed in Supplementary Table~S1). C) Anatomical ROIs considered in the study. Red stripes on the rendered brain indicate the displayed brain sections. *: $p < 0.05$; **: $p < 0.01$; ***: $p < 0.001$; ****: $p < 0.0001$. aPE: absence prediction error. IC: inferior colliculus (auditory midbrain); MGB: medial geniculate body (auditory thalamus); PAC: auditory cortex.
    \label{fig:omres}}
  \end{figure}

\subsection*{Responses to absence in MGB and PAC encode absence prediction error}

\begin{figure}[bth!]
    \centering
    \includegraphics[width=\textwidth]{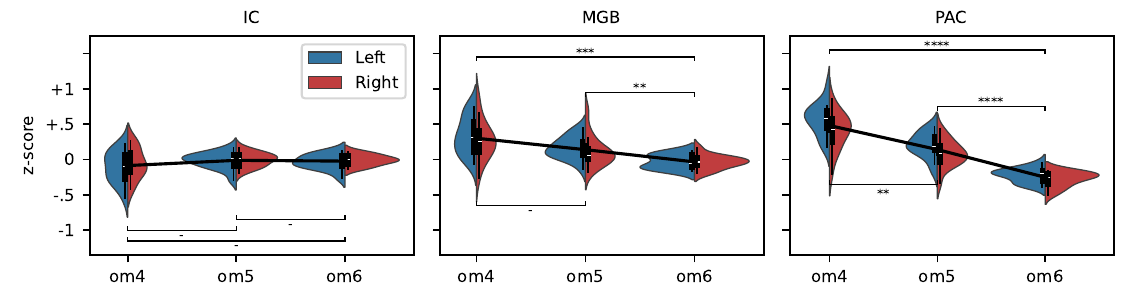}
    \caption{\textbf{Magnitude of the responses to each omission position}. Responses to the omissions ($om4$, $om5$, $om6$) are negatively related to the expectations on the omissions. The effect increases qualitatively along the ascending auditory pathway (for formal quantitative testing see Figure~\ref{fig:emergence}). Responses to the remaining experimental conditions are shown in Supplementary Figure~S2. Blue halves of the violin plots correspond to the left hemisphere; red halves to the right hemisphere. Violin plots are kernel density estimations across the entire population (one sample per participant). All $p$-values were computed using one-sided ranksum tests and Holm-Bonferroni corrected for 15 comparisons (nominal corrected and uncorrected $p$-values are listed in Supplementary Table~S2). *: $p < 0.05$; **: $p < 0.01$; ***: $p < 0.001$; ****: $p < 0.0001$. aPE: absence prediction error. IC: inferior colliculus (auditory midbrain); MGB: medial geniculate body (auditory thalamus); PAC: auditory cortex. $om\,\cdot$: omission in position 4, 5, or 6.
    \label{fig:allregs}}
  \end{figure}

  In agreement with our main hypothesis (the encoding of aPE), the responses to the omissions significantly showed the expected dependency with stimulus presence plausibility in MGB and PAC (Figure~\ref{fig:allregs}; e.g., for the contrast $om6 > om4$: $p<10^{-3}$, $d \geq 1.5$; all effect sizes and $p$-values reported in Supplementary Table~S2). In contrast, there was no detectable aPE in IC (distributions centred around zero, $d \in [-0.7, -0.03]$).

  The results were qualitatively replicated between individual participants (Supplementary Figure S3). Omissions elicited a significant response that was significantly inversely-related to the predictability of the omissions in the ICs of no participants, in the MGBs of 5/15 participants, and in the PACs of 14/15 participants.

\subsection*{Absence prediction error emerges along the ascending auditory pathway}

\begin{figure}[bth!]
    \centering
    \includegraphics[width=\textwidth]{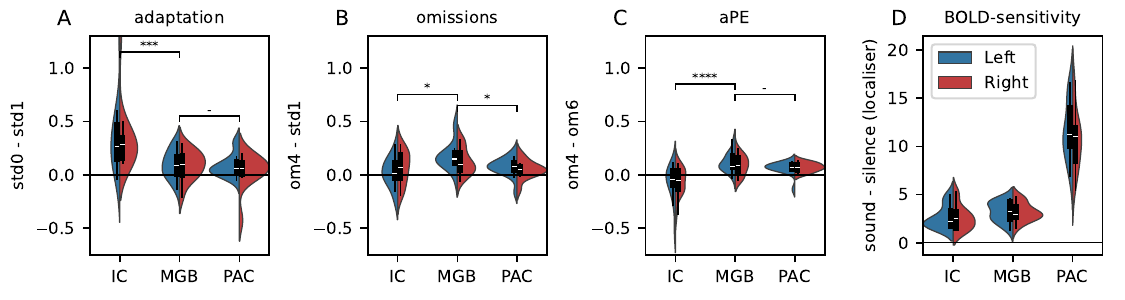}
    \caption{\textbf{Adaptation, omission, and aPE effects normalised by BOLD sensitivity}. A-C) Adaptation (A, $std0 > std1$), omissions (B, $om4 > std1$) and aPE (C, $om4 > om6$) contrasts, voxel-wise normalised by the BOLD sensitivity contrast, and averaged across all voxels. D) BOLD sensitivity as measured by the average contrast $sound > silence$ of the independent functional localiser in each voxel. Violin plots are kernel density estimations across the entire population (one sample per participant). Blue halves of the violin plots correspond to the left hemisphere; red halves to the right hemisphere. All $p$-values were computed using one-sided ranksum tests and Holm-Bonferroni corrected for 6 comparisons (nominal corrected and uncorrected $p$-values are listed in Supplementary Table~S3). *: $p < 0.05$; **: $p < 0.01$; ***: $p < 0.001$; ****: $p < 0.0001$. aPE: absence prediction error. IC: inferior colliculus (auditory midbrain); MGB: medial geniculate body (auditory thalamus); PAC: auditory cortex.
    \label{fig:emergence}}
  \end{figure}

  Adaptation, omission responses, and aPE appeared to be expressed in different amounts along the ascending auditory pathway (Figure~\ref{fig:allregs}). This might reflect that aPE emerges along the ascending auditory pathway; alternatively, the effect could be caused by differences in BOLD sensitivity in the different ROIs. To differentiate between these two interpretations and formally test whether aPE emerges along the ascending auditory pathway, we normalised the adaptation, omission, and aPE contrasts by BOLD sensitivity (Figure~\ref{fig:emergence}) and noise levels (Supplementary Figure~S4) of each ROI.

  Whereas the normalised adaptation was significantly stronger in IC than in MGB or PAC (Figure~\ref{fig:emergence}A; $d = 0.89, p = 0.01$), the normalised omission (Figure~\ref{fig:emergence}B; $d = -0.99, p = 0.012$) and aPE contrasts (Figure~\ref{fig:emergence}C; $d = -1.6, p = 3 \times 10^{-6}$) were significantly stronger in MGB than in IC. Critically, the normalised distributions for aPE were centred around zero in bilateral IC indicating no response. The differences in the adaptation ($d = -0.04$), omission ($d = 0.33$) and aPE ($d = -0.22$) contrasts were not significantly different between MGB and PAC after ($p \geq 0.47$) or before ($p \geq 0.16$) correction for multiple comparisons.

  Similar results were obtained for contrasts normalised by noise levels rather than by BOLD sensitivity (Supplementary Figure~S4 and Supplementary Table~S3), confirming that aPE emerges in MGB and does not seem to increase towards PAC.

  \section*{Discussion} 
    

  Our main aim was to test whether the brain uses a dedicated type of prediction error to update its internal beliefs on the absence of a stimulus. We used the human auditory pathway as a model for sensory processing. There were three key results: (i) Unexpected absence (omissions) elicited activation in IC (auditory midbrain), MGB (auditory thalamus), and PAC (primary auditory cortex), even in the absence of sensory inputs. This finding constitutes the first demonstration of omission responses in the subcortical sensory pathway \cite{Jiang2022, Lao2023}. (ii) Responses elicited by omissions were related to the expectations held by participants on the presence of the omitted tone, demonstrating that they encode absence prediction error. This finding constitutes the first unambiguous demonstration of the encoding of absence prediction error in the brain \cite{Heilbron2017, Braga2022}. (iii) Absence prediction error was encoded in the MGB and PAC, but not in the IC. This pattern is distinct from feature prediction error which is strongly encoded in IC \cite{Tabas2020, Tabas2024, Tabas2024a, Ara2024}. Together, the results suggest that the detection of absence is based on a specific type of prediction error in the human brain and that it is computed by dedicated circuits.

\paragraph{} 

  A central assumption of predictive coding as grounded in the Free Energy Principle \cite{Friston2003, Friston2009} (see \cite{Millidge2022} for a comprehensive review) under the Laplace assumption is that probabilistic representations are Gaussianly distributed. This parametrisation is sufficient to model the processing of an oddball trial, in which the only perceptual object to be inferred is one characteristic of the stimuli such as the fundamental frequency or pitch of the tones. In this case, prior and posterior beliefs can then be modelled with Gaussian distributions centred in the predicted and perceived pitch, respectively. Updating the prior would require only to modify the mean, which can be accomplished with what we call \emph{feature prediction error}. The remaining adjustments in the representation, including the removal of the peak around the pitch of the standard, can be deduced from the parametric constraints of the representation.
  
  However, most perceptual problems require more flexible probabilistic representations \cite{Aitchison2017, Koblinger2021a}. For instance, our omission paradigm cannot be modelled using Gaussian representations: what would be the mean and variance representing an omitted tone? How would such representation extrapolate to the presentation not of one tone, but two simultaneous tones with different pitch? Feature-based prediction error is not sufficient to update internal beliefs on the sensory world: the brain needs a way to detect whether the different elements of the sensory world are present or absent. We propose that the \emph{absence prediction error} introduced in this work is paramount to this process.

\paragraph{} 

  Feature prediction errors can be positive and negative \cite{Keller2018, Hertag2025}. Positive and negative errors are thought to have opposite functional roles that require opposite physiological properties: while positive error signals serve to increase the value of the feature in an internal representation, negative error signals serve to decrease it. We speculate that this separation between positive and negative might also be a hallmark of the error signals responsible for the update of beliefs on occurrence introduced in this work: while absence prediction error might be specifically dedicated to decreasing the belief that a stimulus has occurred, a \emph{presence} counterpart might be specifically dedicated to increasing it.

  One interpretation of the present results is that the IC encodes only presence prediction error, while MGB and PAC encode both, presence and absence prediction error. Another interpretation is that, whereas IC encodes feature prediction error, MGB and PAC encode both, presence/absence and feature PE.

\paragraph{} 
  
  Prediction error has been extensively studied \cite{Heilbron2017, Walsh2020}. Although most previous work has focused on the cerebral cortex \cite{Heilbron2017, Walsh2020}, extensive work has demonstrated the encoding of prediction error also subcortically in the rodent \cite{Parras2017, Lesicko2022}, primate \cite{Jiang2022}, and human \cite{Cacciaglia2015, Tabas2020, Tabas2024a, Tabas2024, Ara2024} auditory pathways \cite{Parras2017}. Whereas several studies have found feature prediction error in the human brain to be equally strong in the IC as it is in the MGB \cite{Tabas2020, Tabas2024a, Tabas2024, Ara2024}, one study on rodents \cite{Parras2017} reported that prediction error emerges along the auditory pathway. This divergence between the human and rodent studies might be a consequence of the use of different species and paradigms.

\paragraph{} 

  Responses to omissions have been robustly shown in primary sensory and association areas of the cerebral cortex in humans, primates, and rodents (mostly in the auditory modality \cite{Braga2022}, but also recently in primary visual \cite{Nitzan2025} and secondary somatosensory \cite{Andersen2019} cortices). Responses to unexpected omissions have not been shown robustly before in a subcortical sensory pathway, although there have been recent attempts in the rodent IC \cite{Lao2023}. Omission responses in omission paradigms can be interpreted as a hallmark of prediction within the predictive coding framework \cite{Heilbron2017, Braga2022, Bendixen2012}, but also as a result of neural entrainment to stimulus repetition \cite{Will2007, LHermite2023}. Our paradigm used abstract rules to successfully disambiguate between these two interpretations. Our results illustrate that, whereas omission responses in MGB and PAC are best explained under the predictive coding framework, omission responses in IC are likely the result of neural entrainment to stimulus repetition because, although the right IC weakly responds to omissions, its responses do not scale with predictability.

\paragraph{} 

  Omission responses have been interpreted before as neural correlates of predictions \cite{Heilbron2017, Berlot2018, Walsh2020, Braga2022, Audette2022}. Under the predictive coding framework, a neuronal response elicited in the absence of sensory input could carry out either, the top-down prediction for the tone, or the bottom-up reaction to the unfulfilled prediction \cite{Friston2003, Spratling2017}. In our view, the omission responses reported here could incorporate contributions from both, errors and predictions. Finer temporal resolution (to tell apart the top-down signals that precede the bottom-up reactions) and/or finer spatial resolution (sufficient to differentiate whether signals are transmitted from or to the areas of interest; e.g., by inspecting laminar cortical profiles \cite{Bastos2012}) will be necessary to disambiguate between bottom-up or top-down signals.
  
\paragraph{} 

  Previous work suggested a mechanistic dissociation between the detection of a stimulus and the identification of its features. First, predictions on feature value and on the plausibility of presence affect the perceptual accuracy of each other in asymmetric ways \cite{Haarsma2025}. Second, detecting absence is an intrinsically different process from detecting presence: for the latter, the brain needs to collect direct evidence that a stimulus is there; for the former, the brain depends indirectly on the lack of evidence that the stimulus is there \cite{Merten2012, Mazor2025, Mazor2026}. As a consequence, human observers respond faster when reporting presence than when reporting absence \cite{Mazor2021, Mazor2023a}. This difference does not exist when the brain identifies the features of a stimulus; e.g., the evidence for a tone's pitch is always direct and does not depend on the lack of evidence for a different value. The present study shows that the dissociation suggested by this previous work has a direct neurobiological substrate in the subcortical auditory pathway, where absence and feature prediction errors are computed at different stages of the sensory hierarchy.
  
\paragraph{} 

  Here we have introduced and provided robust evidence for the encoding of \emph{absence prediction error}, a dedicated error signal to determine not the features of a stimulus, but whether a stimulus is present or not. Given the essential role that predictions and prediction errors play in perceptual inference \cite{DeLange2018, Aitchison2017, Friston2003}, an atypical handling of prediction error could have dramatic effects on the way we perceive our surroundings \cite{McFadyen2020, Diaz2012, Tabas2021}. For instance, the hallucinations typical of psychosis can be understood as the failure of perceptual systems to update the plausibility of incorrectly predicted stimuli from the internal representations: i.e., despite the silence, psychotic individuals hear the voices they predict \cite{Sterzer2018, Keller2024}. Understanding the dissociated computation of feature and presence/absence prediction errors in sensory pathways is essential to our understanding of perceptual function and dysfunction.

  \section*{Methods}
    
    
    The study was approved by the Ethics Committee of the Technische Universit\"at Dresden, Germany (ethics approval number SR-EK-293062022). All participants provided their informed written consent and received monetary compensation for their participation. We preregistered the study after the pilot data collection and before the data collection started \cite{Tabas2022prereg}. This report includes only data acquired after preregistration. The results of the pilot data, including 5 participants, are qualitatively identical to those described here.

\subsection*{Participants}

    Data collection was optimised to show effects at the single participant level. We decided to acquire data from 15 independent participants as biological replications to test the generalisability of the results. The 15 participants were German native speakers (10 female), aged 19 to 34 years (mean 26.1). None of them reported a history of psychiatric or neurological disorders, hearing difficulties, or current use of psychoactive medication. Normal hearing abilities were confirmed with pure tone audiometry (250\,Hz to 8000\,Hz); all participants had hearing thresholds equal to or below 15\,dB SPL in the frequency range of the stimuli used in the experiment (1000\,Hz-3000\,Hz). Participants were also screened for developmental dyslexia (German SLRT-II test \cite{Moll2014}, RST-ARR \cite{Ibrahimovic2013}, and rapid automatised naming (RAN) test of letters, numbers, objects, and colours \cite{Denckla1974}) and autism spectrum disorder (Autism Spectrum Quotient; AQ \cite{Baron2001}). All scores were within the typical range (SLRT: all $PR_{\text{words}} \geq  31, PR_{\text{pseudowords}} \geq  16$, higher than or equal to the cut-off value of 16 \cite{Gutschmidt2020}; RST-ARR: all $PR \geq 61$, higher than the cut-off value of 16; RAN: maximum of 1 error and $RT < 22.3$ seconds in the categories letters and numbers; AQ: all participants $AQ \leq 20$, under the cut-off value of 32).

\subsection*{Experimental paradigm}

    \subsubsection*{Main experiment}

        A trial consisted of a sequence of up to eight repetitions of a pure tone separated by a constant inter-stimulus-interval $ISI = 500$\,ms (Figure~\ref{fig:design}). Trials contained one omission (a missing pure tone) with probability $p_0 = 6/7$. Omissions occurred at position 4, 5, or 6 in the sequence. We asked the participants to attend and report, via a button press, the position of the omission within the sequence, if any. All pure tones had a 50\,ms duration and a frequency $f_n$ sampled from a set of 21 log-distributed frequencies $f_n \in [500, 1500]\,$Hz. Tones were ramped in and out by 5\,ms duration Hanning windows.

        Data collection was optimised to show effects at the single participant level by acquiring up to 882 trials per participant. The number of trials per participant was decided after the evaluation of the pilot results of n=5 participants to guarantee a statistical power of $1-\beta > 0.9$ at the level of the single participant. 

        Trials were separated by jittered inter-trial-intervals (ITI) that were calculated so that the omissions were separated by an average of 5 seconds, up to a maximum of 11 seconds, with a minimum $ITI$ of 1500\,ms. We chose such ITI properties to maximise the efficiency of the estimation of the responses \cite{Friston1999}, while keeping a sufficiently long ITI to ensure that the sequences belonging to separate trials were not confused by the participants.

        To keep the number of conditions (i.e., omissions at position 4, 5, and 6) precisely balanced and $p_0 = 6/7$, the first two runs in a session comprised 74 trials (21 per omission position and 11 without an omission) and the last two runs comprised 73 trials (21 per omission position and 10 without omissions). In addition, each run included 22 null events: silent trials with no stimuli or task of the same duration of a trial. Null events were included to optimise the efficiency of the estimation of the responses \cite{Friston1999} and were pseudorandomly distributed along each run with the following constraints: 1) null events did not occur at the beginning or end of the run; 2) no two null-events occurred successively. Each run lasted about 11 minutes. There were four runs per session. Within each session, trials were balanced across omission positions and the frequency of the pure tone so that each of the 21 pure tones was used four times with every omission position and two times in trials without omissions. Trial ordering was pseudorandomised for each session and participant. Each session lasted around an hour and 3 sessions per participant were planned. In total, across the three sessions, each participant completed a total of 882 trials: 252 for each omission position, and 126 trials without omissions. 

        Thirteen of the fifteen participants completed three sessions with four runs in each session; i.e., 12 runs in total. The remaining two participants took part in two sessions; one with eight runs in total, the other with seven runs. For one of the participants, the auditory midbrain was only partially included in the MRI acquisition in one of the sessions; therefore, we excluded the four affected runs. Two additional participants were recruited, but they completed only one session and were therefore excluded from the study. Each session was taken on 3 different days per participant.

    \subsubsection*{Functional localiser}

        We also ran a functional localiser that was designed to elicit responses in the participant's auditory midbrain and auditory thalamus. The localiser contained 10 blocks with 16 contiguous sounds each. In addition, there were 10 blocks with silence. The sounds were taken from a collection of 85 natural sounds assembled by a previous study and had a duration of one second each\cite{Moerel2015}. Participants pressed a key when the same sound was repeated twice, which occurred in 5\% of the trials. The participants received this task to ensure that they attended the sounds. Each functional localiser block lasted 16 seconds and each run lasted about 6.5 minutes. 

\subsection*{Data acquisition}

    Each session of the main experiment included a single run of the functional localiser. The thirteen participants who completed 3 sessions therefore had 3 runs of the functional localiser; the remaining two participants had 2 runs of the functional localiser.
    
    Each of the sessions consisted, in this order, of two runs of the main task, a single volume whole-head echo planar imaging (EPI) during rest, a fieldmap, and, only on the first session, a structural MRI, one more run of the main task, one run of the functional localiser, and a final run of the main task. The session order was arranged to allow participants plenty of rest between runs of the main task to minimise potential fatigue. 

    MRI data were acquired using a Siemens Prisma 3\,T (Siemens Healthineers, Erlangen, Germany) with a 64-channel head coil. Functional MRI data were acquired using EPI sequences. We used  partial coverage with 24 slices, i.e., a slab. The volume was oriented parallel to the superior temporal gyrus such that the slices covered the auditory midbrain, the auditory thalamus, and the superior temporal gyrus. The additional whole-head EPI volume was acquired with the same parameters (including the FoV), but 84 slices to aid the coregistration process (see~\emph{Data preprocessing}). Cardiac signal was acquired using a pulse oximeter (Siemens Healthineers, Erlangen, Germany) throughout each MRI session.

    The EPI sequence had the following acquisition parameters: interleaved slice acquisition; repetition time $TR = 2100$\,ms; echo time $TE = 46$\,ms; flip angle $FA = 68$ degrees; field of view $FOV = 153\times153$\,mm; matrix size = $102\times102$; voxel size$ = 1.5$\,mm isotropic. Structural images were recorded using an MPRAGE \cite{Brant1992} T1 protocol with 1\,mm isotropic resolution and $FOV = 256\times256$\,mm.

    Stimuli were presented using MATLAB (The Mathworks Inc., Natick, MA, USA) with the Psychophysics Toolbox \cite{Brainard1997} and delivered through an Optoacoustics (Optoacoustics Ltd, Or Yehuda, Israel) amplifier and headphones equipped with active noise-cancellation. Loudness was adjusted independently for each participant to a comfortable level before starting data acquisition.

\subsection*{Data preprocessing} 

    Data was preprocessed using fMRIPrep 23.2.2 \cite{Esteban2019, Esteban2018}, which is based on \emph{Nipype} 1.8.6 \cite{Gorgolewski2011, Gorgolewski2018}.

    The T1w image was corrected for intensity non-uniformity (INU) with \emph{N4BiasFieldCorrection} \cite{Tustison2010}, distributed with ANTs 2.5.0 \cite{Avants2008}, and used as T1w-reference throughout the workflow. The T1w-reference was then skull-stripped with ANTs (version 2.5.0), using \emph{OASIS30ANTs} as target template. Brain tissue segmentation of cerebrospinal fluid (CSF), white-matter (WM) and gray-matter (GM) was performed on the brain-extracted T1w using \emph{fast} (FSL) \cite{Zhang2001}. Brain surfaces were reconstructed using \emph{recon-all} (FreeSurfer 7.3.2) \cite{Dale1999}, and the skull-stripping brain mask was refined with a custom variation of the method to reconcile ANTs-derived and FreeSurfer-derived segmentations of the cortical gray-matter of Mindboggle \cite{Klein2017}. Volume-based spatial normalisation to MNI \cite{Fonov2009} was performed by nonlinear registration with ANTs, using brain-extracted versions of both the T1w reference and the MNI template.

    For the preprocessing of the functional data, a reference volume was first generated using a method in \emph{fMRIPrep} for head-motion correction. Head-motion parameters with respect to the BOLD reference (transformation matrices, and six corresponding rotation and translation parameters) were estimated before any spatiotemporal filtering using FSL's \emph{mcflirt} \cite{Jenkinson2002}. The BOLD reference was then co-registered to the T1w reference using FreeSurfer's \emph{bbregister} \cite{Greve2009}, which implements boundary-based registration. Co-registration was configured with six degrees of freedom. 

    Several confounding time-series were calculated based on the \emph{preprocessed BOLD}: framewise displacement (FD), DVARS and two region-wise global signals extracted within the CSF and the WM. FD was computed using two formulations following Power (absolute sum of relative motion \cite{Power2014}) and Jenkinson (relative root mean square displacement between affines \cite{Jenkinson2002}). Frames that exceeded a threshold of 0.5 mm FD or 1.5 standardized DVARS were annotated as motion outliers. 

    All resamplings from the functional space to the participant's T1w space (for participant-level analyses) and to the MNI space (for group-level analyses) were performed with \emph{a single interpolation step} by composing all the pertinent transformations (i.e.~head-motion transform matrices, susceptibility distortion correction when available, and co-registrations to anatomical and output spaces). Gridded (volumetric) resamplings were performed using \emph{nitransforms}, configured with cubic B-spline interpolation.

\subsection*{Estimation of the average BOLD responses}

    First level analyses (estimation of the responses to the main regressors) were coded in Nipype and carried out using SPM12. The first level GLM's design matrix for the main experiment included 6 regressors (Figure~\ref{fig:design}C): first standard tone ($std0$), standard tones before the omission or before the sixth position in trials without omissions ($std1$), standard tones after the omission or after the sixth position in trials without omissions ($std2$), and omissions in positions 4, 5, and 6 ($om4$, $om5$, and $om6$, respectively).

    We assumed that the responses to the repeated standards ($std1$ and $std2$) varied approximately linearly across successive repetitions and therefore modelled the conditions $std1$ and $std2$ using linear parametric modulation \cite{ODoherty2007}. The linear factors were coded according to the position of the sound within the sequence to account for effects of habituation \cite{Tabas2020}. This design allowed us to maximally disentangle responses to stimuli that were close to each other in time.

    The first level GLM's design matrix for the functional localiser included 2 conditions: sound and silence. We used SPM's contrast estimation to compute the $t$-statistics of the contrast $sound > silence$ for each participant independently. The resulting $t$-maps were used to compute the regions of interest (ROIs) for each participant (see section Definition of the ROIs).

    The design matrices also included six motion parameters (three for translation and three for rotation) and two additional data-driven denoising regressors (the average BOLD response in the CSF and in the white matter; see data preprocessing) as nuisance regressors.

\subsection*{Definition of the ROIs}

    We used an anatomical atlas of the subcortical auditory pathway \cite{Sitek2019} to compute masks for the regions of interest (ROIs) corresponding to the left and right auditory midbrain (inferior colliculus; IC) and left and right auditory thalamus (medial geniculate body; MGB), respectively. The atlas comprises three different definitions of the ROIs calculated using 1) data from the big brain project, 2) postmortem data, and 3) fMRI in vivo-data \cite{Sitek2019}. To compute the prior coarse region for each nucleus we combined the three masks and expanded the resulting mask with a Gaussian kernel with $FWHM = 1$\,mm isotropic. This mask was applied to the $t$-maps from the contrasts of the functional localiser. The masked $t$-maps were consequently thresholded to increasingly higher values until the number of surviving voxels equaled the volume of the region reported in \cite{Sitek2019}; namely, 146 voxels for auditory midbrain (each), and 152 for auditory thalamus (each).

    We computed the ROIs for bilateral primary auditory cortex (anterior transverse temporal gyrus of Heschl, as described in the Destrieux atlas \cite{Destrieux2010}; aHG) in the surface space independently for each participant as part of the \emph{recon-all} routine of Freesurfer \cite{Dale1999} via fMRIPrep (see Section Data preprocessing).

\subsection*{Statistical analyses}

    Beta values encoding the average response to each experimental condition computed with the GLMs were first $z$-scored per run and participant before running statistics to ensure they all had zero mean and unit variance. 

    Analyses at the single-participant level used the estimates for each run as an independent sample (so one point per run) and beta estimates in the T1 space of the participant. Analyses at the group level used the average estimate across run per participant as an independent sample (so one point per participant) and beta estimates in MNI space.

    We used non-parametric ranksum tests for all pair-wise comparisons and Holm-Bonferroni corrected all $p$-values for multiple comparisons. The exact number of comparisons is listed for each specific test in the Result section. We did not correct for multiple comparisons across the 15 participants because we treat the participants as independent biological replications of the main findings; however, we corrected for multiple comparisons across ROIs and pairwise comparisons since all the analyses for a single participant are part of the main hypothesis. 

    We considered a result significant if the corrected $p$-value was lower than $\alpha = 0.05$. We used custom python scripts for all statistical analyses over the beta estimates (see Section Data and code availability).

 \subsection*{Differences to preregistration plans}

     We deviated from our preregistered plans \cite{Tabas2022prereg} in four ways: First, rather than using a custom pipeline for data preprocessing (as anticipated in the preregistration), we decided to use standardised fMRIPrep \cite{Esteban2019}. When we wrote the preregistration, we did not find the preprocessing of partial brain coverage by fMRIPrep of sufficient quality; however, the current version of fMRIPrep handles partial brain coverage well and incorporates more advanced and updated tools than our original pipeline. Second, due to a technical problem in the recording of physiological responses introduced during an upgrade of our MRI machine, we were unable to use the participants' heart rate to generate physiological nuisance regressors. Instead, we used two data-driven time series: the global average signal in cerebrospinal fluid (CSF) and white matter (WM). Third, we originally planned to estimate responses to the sixth standard in trials with no omissions as part of the analysis. However, after data collection, we realised that the number of trials we collected for that condition ($N = 10.5$ per run on average) was not enough for reliable estimation at our current tSNR. Therefore, we integrated that standard in the $std2$ regressor (Figure~S1). Last, we performed an additional analysis, which we did not anticipate, where we normalised the effect sizes by the BOLD sensitivity of each region. We only realised later in the project that different ROIs displayed dramatically different BOLD sensitivities (Figure~\ref{fig:emergence}) and wanted to corroborate that the absence of the encoding of absence prediction error in auditory midbrain is not driven by a relatively lower sensitivity.

  \subsection*{Data and code availability}

    Logfiles and beta estimates (\url{https://zenodo.org/records/17603000}) as well as all the code used to present the paradigm, analyse the data, and produce each of the figures and tables of this paper (\url{https://github.com/qtabas/omres}) are publicly available. Raw data is not publicly available because we do not have permission from the participants to publish their personal data (high-resolution functional and structural images).

  \section*{Acknowledgements}
    This work was funded by the ERC (Sensocom 647051, granted to KvK; SynPrePro 101115798, granted to AT). The MAGNETOM Prisma upgrade was funded by the German Research Foundation (DFG; INST 269/881-1 FUGG).

  \bibliographystyle{ieeetr}
  \bibliography{bib}

@article{Aitchison2017,
	author = {Aitchison, Laurence and Lengyel, M{\'{a}}t{\'{e}}},
	doi = {10.1016/j.conb.2017.08.010},
	issn = {18736882},
	journal = {Current Opinion in Neurobiology},
	pages = {219--227},
	pmid = {28942084},
	title = {{With or without you: predictive coding and Bayesian inference in the brain}},
	volume = {46},
	year = {2017}
}

@article{Andersen2019,
	title = {Somatosensory responses to nothing: {An} {MEG} study of expectations during omission of tactile stimulations},
	volume = {184},
	issn = {1053-8119},
	doi = {10.1016/j.neuroimage.2018.09.014},
	language = {en},
	urldate = {2025-07-10},
	journal = {NeuroImage},
	author = {Andersen, Lau M. and Lundqvist, Daniel},
	year = {2019},
	note = {Publisher: Elsevier BV},
	pages = {78--89},
}

@article{Antunes2021,
	author = {Flora M. Antunes and Manuel S. Malmierca},
	doi = {10.3389/fncir.2021.721186},
	issn = {16625110},
	journal = {Frontiers in Neural Circuits},
	month = {8},
	pmid = {34489648},
	publisher = {Frontiers Media S.A.},
	title = {Corticothalamic Pathways in Auditory Processing: Recent Advances and Insights From Other Sensory Systems},
	volume = {15},
	year = {2021},
}

@article{Ara2024,
	title = {Cortical--Subcortical Interactions Underlie Processing of Auditory Predictions Measured with {{7T fMRI}}},
	author = {Ara, Alberto and Provias, Vasiliki and Sitek, Kevin and Coffey, Emily B J and Zatorre, Robert J},
	year = {2024},
	month = aug,
	journal = {Cerebral Cortex},
	volume = {34},
	number = {8},
	pages = {bhae316},
	issn = {1047-3211, 1460-2199},
	doi = {10.1093/cercor/bhae316},
	urldate = {2025-04-01},
}

@article{Avants2008,
	author = {Avants, B.B. and Epstein, C.L. and Grossman, M. and Gee, J.C.},
	doi = {10.1016/j.media.2007.06.004},
	issn = {1361-8415},
	journal = {Medical Image Analysis},
	number = 1,
	pages = {26-41},
	shorttitle = {Symmetric diffeomorphic image registration with cross-correlation},
	title = {Symmetric diffeomorphic image registration with cross-correlation: Evaluating automated labeling of elderly and neurodegenerative brain},
	url = {https://www.sciencedirect.com/science/article/pii/S1361841507000606},
	volume = 12,
	year = 2008
}

@article{Audette2022,
	title = {Precise Movement-Based Predictions in the Mouse Auditory Cortex},
	author = {Audette, Nicholas J. and Zhou, WenXi and La Chioma, Alessandro and Schneider, David M.},
	year = {2022},
	journal = {Current Biology},
	volume = {32},
	number = {22},
	pages = {4925-4940.e6},
	issn = {09609822},
	doi = {10.1016/j.cub.2022.09.064},
}

@article{Baron2001,
	author = {Baron-Cohen, S and Wheelwright, S and Hill, J and Raste, Y and Plumb, I},
	doi = {10.1017/S0021963001006643},
	issn = {00219630},
	journal = {Journal of Child Psychology and Psychiatry and Allied Disciplines},
	number = {2},
	pages = {241--251},
	title = {{The "Reading the Mind in the Eyes" Test revised version: A study with normal adults, and adults with Asperger syndrome or high-functioning autism}},
	volume = {42},
	year = {2001}
}

@article{Bastos2012,
	author = {Bastos, Andre M and Usrey, W Martin and Adams, Rick a and Mangun, George R and Fries, Pascal and Friston, Karl J},
	doi = {10.1016/j.neuron.2012.10.038},
	issn = {1097-4199},
	journal = {Neuron},
	month = {nov},
	number = {4},
	pages = {695--711},
	pmid = {23177956},
	publisher = {Elsevier Inc.},
	title = {{Canonical microcircuits for predictive coding.}},
	volume = {76},
	year = {2012}
}

@article{Beck2014,
	author = {Claudia Beck and Bernd Kardatzki and Thomas Ethofer},
	doi = {10.1371/journal.pone.0084667},
	issn = {19326203},
	issue = {1},
	journal = {PLoS ONE},
	pmid = {24416261},
	title = {Mondegreens and Soramimi as a method to induce misperceptions of speech content - Influence of familiarity, wittiness, and language competence},
	volume = {9},
	year = {2014},
}

@article{Bendixen2012,
	author = {Bendixen, Alexandra and SanMiguel, Iria and Schr{\"{o}}ger, Erich},
	doi = {10.1016/j.ijpsycho.2011.08.003},
	issn = {01678760},
	journal = {International Journal of Psychophysiology},
	number = {2},
	pages = {120--131},
	publisher = {Elsevier B.V.},
	title = {{Early electrophysiological indicators for predictive processing in audition: A review}},
	volume = {83},
	year = {2012}
}

@article{Berlot2018,
	author = {Berlot, Eva and Formisano, Elia and {De Martino}, Federico},
	doi = {10.1523/JNEUROSCI.2205-17.2018},
	issn = {15292401},
	journal = {Journal of Neuroscience},
	number = {21},
	pages = {4934--4942},
	pmid = {29712781},
	title = {{Mapping frequency-specific tone predictions in the human auditory cortex at high spatial resolution}},
	volume = {38},
	year = {2018}
}

@article{Braga2022,
	author = {Braga, Alessandro and Sch{\"{o}}nwiesner, Marc},
	doi = {10.3389/fncir.2022.799581},
	issn = {16625110},
	journal = {Frontiers in Neural Circuits},
	number = {February},
	pages = {1--14},
	pmid = {35177967},
	title = {{Neural Substrates and Models of Omission Responses and Predictive Processes}},
	volume = {16},
	year = {2022}
}

@article{Brainard1997,
	author = {Brainard, David H.},
	doi = {10.1163/156856897X00357},
	issn = {01691015},
	journal = {Spatial Vision},
	number = {4},
	pages = {433--436},
	pmid = {9176952},
	title = {{The Psychophysics Toolbox}},
	volume = {10},
	year = {1997}
}

@article{Brant1992,
	author = {Brant-Zawadzki, Michael and Gillan, Gary D. and Nitz, Wolfgang R.},
	doi = {10.1148/radiology.182.3.1535892},
	issn = {00338419},
	journal = {Radiology},
	number = {3},
	pages = {769--775},
	pmid = {1535892},
	title = {{MP RAGE: A three-dimensional, T1-weighted, gradient-echo sequence - Initial experience in the brain}},
	volume = {182},
	year = {1992}
}

@article{Cacciaglia2015,
	author = {Cacciaglia, Raffaele and Escera, Carles and Slabu, Lavinia and Grimm, Sabine and Sanju{\'{a}}n, Ana and Ventura-Campos, Noelia and {\'{A}}vila, C{\'{e}}sar},
	doi = {10.1016/j.neuropsychologia.2015.01.001},
	issn = {18733514},
	journal = {Neuropsychologia},
	pages = {51--58},
	pmid = {25556848},
	publisher = {Elsevier},
	title = {{Involvement of the human midbrain and thalamus in auditory deviance detection}},
	volume = {68},
	year = {2015}
}

@article{Carbajal2018,
	author = {Carbajal, Guillermo V and Malmierca, Manuel S},
	doi = {10.1177/2331216518784822},
	issn = {2331-2165},
	journal = {Trends in Hearing},
	pages = {233121651878482},
	title = {{The Neuronal Basis of Predictive Coding Along the Auditory Pathway: From the Subcortical Roots to Cortical Deviance Detection}},
	volume = {22},
	year = {2018}
}

@article{Chandrasekaran2009,
	author = {Chandrasekaran, Bharath and Hornickel, Jane and Skoe, Erika and Nicol, Trent and Kraus, Nina},
	doi = {10.1016/j.neuron.2009.10.006},
	isbn = {1097-4199 (Electronic)$\backslash$r0896-6273 (Linking)},
	issn = {08966273},
	journal = {Neuron},
	number = {3},
	pages = {311--319},
	pmid = {19914180},
	publisher = {Elsevier Ltd},
	title = {{Context-Dependent Encoding in the Human Auditory Brainstem Relates to Hearing Speech in Noise: Implications for Developmental Dyslexia}},
	volume = {64},
	year = {2009}
}

@article{Dale1999,
	author = {Dale, Anders M. and Fischl, Bruce and Sereno, Martin I.},
	doi = {10.1006/nimg.1998.0395},
	issn = {1053-8119},
	journal = {NeuroImage},
	number = 2,
	pages = {179-194},
	shorttitle = {Cortical Surface-Based Analysis},
	title = {Cortical Surface-Based Analysis: I. Segmentation and Surface Reconstruction},
	url = {https://www.sciencedirect.com/science/article/pii/S1053811998903950},
	volume = 9,
	year = 1999
}

@article{DeLange2018,
	author = {de Lange, Floris P. and Heilbron, Micha and Kok, Peter},
	doi = {10.1016/j.tics.2018.06.002},
	issn = {1879307X},
	journal = {Trends in Cognitive Sciences},
	number = {9},
	pages = {764--779},
	publisher = {Elsevier Ltd},
	title = {{How Do Expectations Shape Perception?}},
	volume = {22},
	year = {2018}
}

@article{Denckla1974,
	author = {Denckla, Martha Bridge and Rudel, Rita},
	doi = {10.1016/S0010-9452(74)80009-2},
	issn = {00109452},
	journal = {Cortex},
	number = {2},
	pages = {186--202},
	title = {{Rapid “Automatized” Naming of Pictured Objects, Colors, Letters and Numbers by Normal Children}},
	volume = {10},
	year = {1974}
}

@article{Destrieux2010,
	title = {Automatic Parcellation of Human Cortical Gyri and Sulci Using Standard Anatomical Nomenclature},
	author = {Destrieux, Christophe and Fischl, Bruce and Dale, Anders and Halgren, Eric},
	year = {2010},
	journal = {NeuroImage},
	volume = {53},
	number = {1},
	pages = {1--15},
	issn = {10538119},
	doi = {10.1016/j.neuroimage.2010.06.010},
}

@article{Diaz2012,
	author = {Diaz, B. and Hintz, F. and Kiebel, S. J. and von Kriegstein, K.},
	doi = {10.1073/pnas.1119828109},
	isbn = {1091-6490},
	issn = {0027-8424},
	journal = {Proceedings of the National Academy of Sciences},
	number = {34},
	pages = {13841--13846},
	pmid = {22869724},
	title = {{Dysfunction of the auditory thalamus in developmental dyslexia}},
	volume = {109},
	year = {2012}
}

@article{Esteban2019,
	author = {Esteban, Oscar and Markiewicz, Christopher and Blair, Ross W and Moodie, Craig and Isik, Ayse Ilkay and Erramuzpe Aliaga, Asier and Kent, James and Goncalves, Mathias and DuPre, Elizabeth and Snyder, Madeleine and Oya, Hiroyuki and Ghosh, Satrajit and Wright, Jessey and Durnez, Joke and Poldrack, Russell and Gorgolewski, Krzysztof Jacek},
	title = {{fMRIPrep}: a robust preprocessing pipeline for functional {MRI}},
	volume = {16},
	pages = {111--116},
	year = {2019},
	doi = {10.1038/s41592-018-0235-4},
	journal = {Nature Methods}
}

@article{Esteban2018,
	author = {Esteban, Oscar and Blair, Ross and Markiewicz, Christopher J. and Berleant, Shoshana L. and Moodie, Craig and Ma, Feilong and Isik, Ayse Ilkay and Erramuzpe, Asier and Kent, James D. andGoncalves, Mathias and DuPre, Elizabeth and Sitek, Kevin R. and Gomez, Daniel E. P. and Lurie, Daniel J. and Ye, Zhifang and Poldrack, Russell A. and Gorgolewski, Krzysztof J.},
	title = {fMRIPrep 23.2.2},
	year = 2018,
	doi = {10.5281/zenodo.852659},
	publisher = {Zenodo},
	journal = {Software}
}

@article{Fonov2009,
	title = {Unbiased nonlinear average age-appropriate brain templates from birth to adulthood},
	author = {Fonov, VS and Evans, AC and McKinstry, RC and Almli, CR and Collins, DL},
	doi = {10.1016/S1053-8119(09)70884-5},
	journal = {NeuroImage},
	pages = {S102},
	volume = {47, Supplement 1},
	year = 2009
}

@article{Font2020,
	author = {Font-Alaminos, Marta and Ribas-Prats, Teresa and Gorina-Careta, Nat{\`{a}}lia and Escera, Carles},
	doi = {10.1016/j.heares.2020.107954},
	issn = {03785955},
	journal = {Hearing Research},
	month = {jan},
	pages = {107954},
	title = {{Emergence of prediction error along the human auditory hierarchy}},
	volume = {399},
	year = {2021}
}

@article{Ford2012,
	author = {Judith M. Ford and Daniel H. Mathalon},
	doi = {10.1016/j.ijpsycho.2011.09.004},
	issn = {01678760},
	issue = {2},
	journal = {International Journal of Psychophysiology},
	month = {2},
	pages = {232-239},
	pmid = {21959054},
	title = {Anticipating the future: Automatic prediction failures in schizophrenia},
	volume = {83},
	year = {2012},
}

@article{Friston1999,
	author = {Friston, K.J. and Zarahn, E. and Josephs, O. and Henson, R.N.A. and Dale, A.M.},
	doi = {10.1006/nimg.1999.0498},
	isbn = {1053-8119 (Print)\n1053-8119 (Linking)},
	issn = {10538119},
	journal = {NeuroImage},
	number = {5},
	pages = {607--619},
	pmid = {10547338},
	title = {{Stochastic Designs in Event-Related fMRI}},
	volume = {10},
	year = {1999}
}

@article{Friston2003,
	author = {Friston, Karl},
	doi = {10.1016/j.neunet.2003.06.005},
	issn = {0893-6080},
	journal = {Neural networks},
	month = {nov},
	number = {9},
	pages = {1325--52},
	pmid = {14622888},
	title = {{Learning and inference in the brain.}},
	volume = {16},
	year = {2003}
}

@article{Friston2009,
	author = {Friston, Karl and Kiebel, Stefan},
	doi = {10.1098/rstb.2008.0300},
	isbn = {0962-8436},
	issn = {1471-2970},
	journal = {Philosophical transactions of the Royal Society of London. Series B, Biological sciences},
	number = {1521},
	pages = {1211--21},
	pmid = {19528002},
	title = {{Predictive coding under the free-energy principle.}},
	volume = {364},
	year = {2009}
}

@article{Gorgolewski2011,
	author = {Gorgolewski, Krzysztof and Burns, Christopher D and Madison, Cindee and Clark, Dav and Halchenko, Yaroslav O and Waskom, Michael L and Ghosh, Satrajit S},
	doi = {10.3389/fninf.2011.00013},
	journal = {Frontiers in Neuroinformatics},
	title = {{Nipype: A Flexible, Lightweight and Extensible Neuroimaging Data Processing Framework in Python}},
	volume = {5},
	year = {2011}
}

@article{Gorgolewski2018,
	author = {Gorgolewski, Krzysztof J. and Esteban, Oscar and Markiewicz, Christopher J. and Ziegler, Erik and Ellis, David Gage and Notter, Michael Philipp and Jarecka, Dorota and Johnson, Hans and Burns, Christopher and Manhães-Savio, Alexandre and Hamalainen, Carlo and Yvernault, Benjamin and Salo, Taylor and Jordan, Kesshi and Goncalves, Mathias and Waskom, Michael and Clark, Daniel and Wong, Jason and Loney, Fred and Modat, Marc and Dewey, Blake E and Madison, Cindee and Visconti di Oleggio Castello, Matteo and Clark, Michael G. and Dayan, Michael and Clark, Dav and Keshavan, Anisha and Pinsard, Basile and Gramfort, Alexandre and Berleant, Shoshana and Nielson, Dylan M. and Bougacha, Salma and Varoquaux, Gael and Cipollini, Ben and Markello, Ross and Rokem, Ariel and Moloney, Brendan and Halchenko, Yaroslav O. and Wassermann , Demian and Hanke, Michael and Horea, Christian and Kaczmarzyk, Jakub and Gilles de Hollander and DuPre, Elizabeth and Gillman, Ashley and Mordom, David and Buchanan, Colin and Tungaraza, Rosalia and Pauli, Wolfgang M. and Iqbal, Shariq and Sikka, Sharad and Mancini, Matteo and Schwartz, Yannick and Malone, Ian B. and Dubois, Mathieu and Frohlich, Caroline and Welch, David and Forbes, Jessica and Kent, James and Watanabe, Aimi and Cumba, Chad and Huntenburg, Julia M. and Kastman, Erik and Nichols, B. Nolan and Eshaghi, Arman and Ginsburg, Daniel and Schaefer, Alexander and Acland, Benjamin and Giavasis, Steven and Kleesiek, Jens and Erickson, Drew and Küttner, René and Haselgrove, Christian and Correa, Carlos and Ghayoor, Ali and Liem, Franz and Millman, Jarrod and Haehn, Daniel and Lai, Jeff and Zhou, Dale and Blair, Ross and Glatard, Tristan and Renfro, Mandy and Liu, Siqi and Kahn, Ari E. and Pérez-García, Fernando and Triplett, William and Lampe, Leonie and Stadler, Jörg and Kong, Xiang-Zhen and Hallquist, Michael and Chetverikov, Andrey and Salvatore, John and Park, Anne and Poldrack, Russell and Craddock, R. Cameron and Inati, Souheil and Hinds, Oliver and Cooper, Gavin and Perkins, L. Nathan and Marina, Ana and Mattfeld, Aaron and Noel, Maxime and Lukas Snoek and Matsubara, K and Cheung, Brian and Rothmei, Simon and Urchs, Sebastian and Durnez, Joke and Mertz, Fred and Geisler, Daniel and Floren, Andrew and Gerhard, Stephan and Sharp, Paul and Molina-Romero, Miguel and Weinstein, Alejandro and Broderick, William and Saase, Victor and Andberg, Sami Kristian and Harms, Robbert and Schlamp, Kai and Arias, Jaime and Papadopoulos Orfanos, Dimitri and Tarbert, Claire and Tambini, Arielle and De La Vega, Alejandro and Nickson, Thomas and Brett, Matthew and Falkiewicz, Marcel and Podranski, Kornelius and Linkersdörfer, Janosch and Flandin, Guillaume and Ort, Eduard and Shachnev, Dmitry and McNamee, Daniel and Davison, Andrew and Varada, Jan and Schwabacher, Isaac and Pellman, John and Perez-Guevara, Martin and Khanuja, Ranjeet and Pannetier, Nicolas and McDermottroe, Conor and Ghosh, Satrajit},
	title = {Nipype},
	year = 2018,
	doi = {10.5281/zenodo.596855},
	publisher = {Zenodo},
	journal = {Software}
}

@article{Greve2009,
	author = {Greve, Douglas N and Fischl, Bruce},
	doi = {10.1016/j.neuroimage.2009.06.060},
	issn = {1095-9572},
	journal = {NeuroImage},
	number = 1,
	pages = {63-72},
	title = {Accurate and robust brain image alignment using boundary-based registration},
	volume = 48,
	year = 2009
}

@article{Gutschmidt2020,
	author = {Gutschmidt, K. and Wenninger, S. and Montagnese, F. and Schoser, B.},
	doi = {10.1007/s00415-020-10161-6},
	isbn = {0123456789},
	issn = {14321459},
	journal = {Journal of Neurology},
	title = {{Dyslexia and cognitive impairment in adult patients with myotonic dystrophy type 1: a clinical prospective analysis}},
	year = {2020}
}

@article{Haarsma2025,
	title={Expectations about presence enhance the influence of content-specific expectations on low-level orientation judgements},
	author={Haarsma, Joost and Kaltenmaier, Aaron and Fleming, Stephen M and Kok, Peter},
	journal={Cognition},
	volume={254},
	pages={105961},
	year={2025}, 
	publisher={Elsevier}
}

@article{Heilbron2017,
	author = {Heilbron, Micha and Chait, Maria},
	doi = {10.1016/j.neuroscience.2017.07.061},
	issn = {18737544},
	journal = {Neuroscience},
	publisher = {The Author(s)},
	title = {{Great expectations: Is there evidence for predictive coding in auditory cortex?}},
	year = {2017}
}

@article{Hertag2025,
	title = {Uncertainty Estimation with Prediction-Error Circuits},
	author = {Hert{\"a}g, Loreen and Wilmes, Katharina A. and Clopath, Claudia},
	year = {2025},
	journal = {Nature Communications},
	volume = {16},
	number = {1},
	pages = {3036},
	issn = {2041-1723},
	doi = {10.1038/s41467-025-58311-6},
}

@book{Ibrahimovic2013,
	title={Rechtschreibtest RST-ARR: aktuelle Rechtschreibregelung: L{\"u}ckendiktate},
	author={Ibrahimovi{\'c}, Nedina and Bulheller, Stephan},
	year={2013},
	publisher={Pearson Assessment \& Information}
}

@article{Jenkinson2002,
	author = {Jenkinson, Mark and Bannister, Peter and Brady, Michael and Smith, Stephen},
	doi = {10.1006/nimg.2002.1132},
	issn = {1053-8119},
	journal = {NeuroImage},
	number = 2,
	pages = {825-841},
	title = {Improved Optimization for the Robust and Accurate Linear Registration and Motion Correction of Brain Images},
	url = {https://www.sciencedirect.com/science/article/pii/S1053811902911328},
	volume = 17,
	year = 2002
}

@article{Jiang2022,
	author = {Jiang, Yuwei and Komatsu, Misako and Chen, Yuyan and Xie, Ruoying and Zhang, Kaiwei and Xia, Ying and Gui, Peng and Liang, Zhifeng and Wang, Liping},
	doi = {10.7554/elife.74653},
	issn = {2050084X},
	journal = {eLife},
	pages = {1--21},
	pmid = {35174784},
	title = {{Constructing the hierarchy of predictive auditory sequences in the marmoset brain}},
	volume = {11},
	year = {2022}
}

@article{Jordan2020,
	title = {Opposing {{Influence}} of {{Top-down}} and {{Bottom-up Input}} on {{Excitatory Layer}} 2/3 {{Neurons}} in {{Mouse Primary Visual Cortex}}},
	author = {Jordan, Rebecca and Keller, Georg B.},
	year = {2020},
	journal = {Neuron},
	volume = {108},
	number = {6},
	pages = {1194-1206.e5},
	issn = {08966273},
	doi = {10.1016/j.neuron.2020.09.024},
}

@article{Keller2018,
	author = {Keller, Georg B. and Mrsic-Flogel, Thomas D.},
	doi = {10.1016/j.neuron.2018.10.003},
	issn = {10974199},
	journal = {Neuron},
	number = {2},
	pages = {424--435},
	pmid = {30359606},
	publisher = {Elsevier Inc.},
	title = {{Predictive Processing: A Canonical Cortical Computation}},
	volume = {100},
	year = {2018}
}

@article{Keller2024,
    title = {Predictive {{Processing}}: {{A Circuit Approach}} to {{Psychosis}}},
    shorttitle = {Predictive {{Processing}}},
    author = {Keller, Georg B. and Sterzer, Philipp},
    year = {2024},
    journal = {Annual Review of Neuroscience},
    volume = {47},
    number = {1},
    pages = {85--101},
    issn = {0147-006X, 1545-4126},
    doi = {10.1146/annurev-neuro-100223-121214}
}

@article{Klein2017,
	author = {Klein, Arno and Ghosh, Satrajit S. and Bao, Forrest S. and Giard, Joachim and Häme, Yrjö and Stavsky, Eliezer and Lee, Noah and Rossa, Brian and Reuter, Martin and Neto, Elias Chaibub and Keshavan, Anisha},
	doi = {10.1371/journal.pcbi.1005350},
	issn = {1553-7358},
	journal = {PLOS Computational Biology},
	number = 2,
	pages = {e1005350},
	title = {Mindboggling morphometry of human brains},
	url = {https://journals.plos.org/ploscompbiol/article?id=10.1371/journal.pcbi.1005350},
	volume = 13,
	year = 2017
}

@article{Koblinger2021a,
	title = {Representations of Uncertainty: Where Art Thou?},
	shorttitle = {Representations of Uncertainty},
	author = {Koblinger, {\'A}d{\'a}m and Fiser, J{\'o}zsef and Lengyel, M{\'a}t{\'e}},
	year = {2021},
	journal = {Current Opinion in Behavioral Sciences},
	volume = {38},
	pages = {150--162},
	issn = {23521546},
	doi = {10.1016/j.cobeha.2021.03.009},
	urldate = {2025-10-01},
}

@article{Kok2012,
	author = {Peter Kok and Janneke F.M. Jehee and Floris P. de Lange},
	doi = {10.1016/j.neuron.2012.04.034},
	issn = {08966273},
	journal = {Neuron},
	month = {7},
	pages = {265-270},
	title = {Less Is More: Expectation Sharpens Representations in the Primary Visual Cortex},
	volume = {75},
	year = {2012},
}

@article{Lao2023,
	author = {Ana B. Lao-Rodríguez  and Karol Przewrocki  and David Pérez-González  and Artoghrul Alishbayli  and Evrim Yilmaz  and Manuel S. Malmierca  and Bernhard Englitz },
	title = {Neuronal responses to omitted tones in the auditory brain: A neuronal correlate for predictive coding},
	journal = {Science Advances},
	volume = {9},
	number = {24},
	pages = {eabq8657},
	year = {2023},
	doi = {10.1126/sciadv.abq8657},
	URL = {https://www.science.org/doi/abs/10.1126/sciadv.abq8657},
}

@article{Lesicko2022,
	author = {Alexandria MH Lesicko and Christopher F Angeloni and Jennifer M Blackwell and Mariella De Biasi and Maria N Geffen},
	doi = {10.7554/eLife.73289},
	issn = {2050-084X},
	journal = {eLife},
	month = {3},
	pmid = {35290181},
	publisher = {eLife Sciences Publications Ltd},
	title = {Corticofugal regulation of predictive coding},
	volume = {11},
	year = {2022},
}

@article{Leonardon2025,
	title = {Differential Modulation of Positive and Negative Prediction Errors by Stimulus Variability in the Mouse Posterior Parietal Cortex},
	author = {Leonardon, Benjamin and Kasavica, Sergej and Raltschev, Constanze and Senn, Walter and Wilmes, Katharina A. and Sachidhanandam, Shankar},
	year = {2025},
	journal = {Communications Biology},
	volume = {8},
	number = {1},
	pages = {1397},
	issn = {2399-3642},
	doi = {10.1038/s42003-025-08797-z},
}

@article{LHermite2023,
	title = {Rhythmic {{Entrainment Echoes}} in {{Auditory Perception}}},
	author = {L'Hermite, Sylvain and Zoefel, Benedikt},
	year = {2023},
	journal = {The Journal of Neuroscience},
	volume = {43},
	number = {39},
	pages = {6667--6678},
	issn = {0270-6474, 1529-2401},
	doi = {10.1523/JNEUROSCI.0051-23.2023},
	urldate = {2025-01-12},
}

@article{May2010,
	title = {Mismatch Negativity ({{MMN}}), the Deviance-Elicited Auditory Deflection, Explained},
	author = {May, Patrick J C and Tiitinen, Hannu},
	year = {2010},
	journal = {Psychophysiology},
	volume = {47},
	number = {1},
	pages = {66--122},
	issn = {00485772},
	doi = {10.1111/j.1469-8986.2009.00856.x},
	isbn = {0048-5772},
	pmid = {19686538},
}

@article{Mazor2021,
	title={Stage 2 Registered Report: Metacognitive asymmetries in visual perception},
	author={Mazor, Matan and Moran, Rani and Fleming, Stephen M},
	journal={Neuroscience of Consciousness},
	volume={7},
	number={1},
	pages={niab025},
	year={2021},
	doi={10.1093/nc/niab025},
}

@article{Mazor2023a,
	title={Paradoxical evidence weighting in confidence judgments for detection and discrimination},
	author={Mazor, Matan and Maimon-Mor, Roni O and Charles, Lucie and Fleming, Stephen M},
	journal={Attention, Perception, \& Psychophysics},
	volume={85},
	pages={2356--2385},
	year={2023},
	doi={10.3758/s13414-023-02710-8},
}

@article{Mazor2025,
	title={Inference about absence as a window into the mental self-model},
	author={Mazor, Matan},
	journal={Open Mind},
	volume={9},
	pages={635--651},
	year={2025},
	doi={10.1162/opmi_a_00206},
}

@article{Mazor2026,
	title={Beliefs about perception shape perceptual inference: An ideal observer model of detection.},
	author={Mazor, Matan and Moran, Rani and Press, Clare},
	journal={Psychological Review},
	volume={133},
	number={2},
	pages={271},
	year={2026},
	publisher={American Psychological Association}
}

@article{Merten2012,
	title={Active encoding of decisions about stimulus absence in primate prefrontal cortex neurons},
	author={Merten, Katharina and Nieder, Andreas},
	journal={Proceedings of the National Academy of Sciences},
	volume={109},
	number={16},
	pages={6289--6294},
	year={2012},
	doi={10.1073/pnas.1121084109},
}

@article{McFadyen2020,
	author = {McFadyen, Jessica and Dolan, Raymond J. and Garrido, Marta I.},
	doi = {10.1038/s41583-020-0287-1},
	issn = {14710048},
	journal = {Nature Reviews Neuroscience},
	number = {5},
	pages = {264--276},
	pmid = {32269315},
	publisher = {Springer US},
	title = {{The influence of subcortical shortcuts on disordered sensory and cognitive processing}},
	volume = {21},
	year = {2020}
}

@article{Millidge2022,
  title = {Predictive {{Coding}}: A {{Theoretical}} and {{Experimental Review}}},
  shorttitle = {Predictive {{Coding}}},
  author = {Millidge, Beren and Seth, Anil and Buckley, Christopher L.},
  year = {2022},
  journal = {arXiv},
  urldate = {2024-06-01},
  archiveprefix = {arXiv},
}

@misc{Moll2014,
	title = {Lese-und Rechtschreibtest (SLRT-II). Weiterentwicklung des Salzburger Lese-und Rechtschreibtests (SLRT), 2., korrigierte Auflage mit erweiterten Normen},
	author = {Moll, K and Landerl, K},
	year = {2014},
	publisher = {Verlag Hans Huber, Bern}
}

@article{Moerel2015,
	author = {Moerel, Michelle and {De Martino}, Federico and Uğurbil, K{\^{a}}mil and Yacoub, Essa and Formisano, Elia},
	doi = {10.1038/srep17048},
	issn = {2045-2322},
	journal = {Scientific reports},
	pages = {17048},
	pmid = {26597173},
	publisher = {Nature Publishing Group},
	title = {{Processing of frequency and location in human subcortical auditory structures.}},
	volume = {5},
	year = {2015}
}

@article{Nitzan2025,
	title = {Diversity of omission responses to visual images across brain-wide regions},
	volume = {11},
	language = {en},
	journal = {Science Advances},
	author = {Nitzan, Noam and Buzsáki, György},
	year = {2025},
	pages = {eadv5651},
}

@article{ODoherty2007,
	author = {O'Doherty, John P. and Hampton, Alan and Kim, Hackjin},
	doi = {10.1196/annals.1390.022},
	isbn = {1573316741},
	issn = {17496632},
	journal = {Annals of the New York Academy of Sciences},
	pages = {35--53},
	title = {{Model-based fMRI and its application to reward learning and decision making}},
	volume = {1104},
	year = {2007}
}

@article{Parras2017,
	author = {Parras, Gloria G. and Nieto-Diego, Javier and Carbajal, Guillermo V. and Vald{\'{e}}s-Baizabal, Catalina and Escera, Carles and Malmierca, Manuel S.},
	doi = {10.1038/s41467-017-02038-6},
	issn = {2041-1723},
	journal = {Nature Communications},
	month = {dec},
	number = {1},
	pages = {2148},
	publisher = {Springer US},
	title = {{Neurons along the auditory pathway exhibit a hierarchical organization of prediction error}},
	volume = {8},
	year = {2017}
}

@article{Power2014,
	author = {Power, Jonathan D. and Mitra, Anish and Laumann, Timothy O. and Snyder, Abraham Z. and Schlaggar, Bradley L. and Petersen, Steven E.},
	doi = {10.1016/j.neuroimage.2013.08.048},
	issn = {1053-8119},
	journal = {NeuroImage},
	number = {Supplement C},
	pages = {320-341},
	title = {Methods to detect, characterize, and remove motion artifact in resting state fMRI},
	url = {https://www.sciencedirect.com/science/article/pii/S1053811913009117},
	volume = 84,
	year = 2014
}

@article{Rao1999,
	author = {Rao, Rajesh P N and Ballard, Dana H},
	doi = {10.1038/4580},
	eprint = {NIHMS150003},
	isbn = {1097-6256},
	issn = {1097-6256},
	journal = {Nature Neuroscience},
	month = {jan},
	number = {1},
	pages = {79--87},
	pmid = {10195184},
	title = {{Predictive coding in the visual cortex: a functional interpretation of some extra-classical receptive-field effects}},
	volume = {2},
	year = {1999}
}

@article{Sitek2019,
	author = {Sitek, Kevin R and Gulban, Omer Faruk and Calabrese, Evan and Johnson, G Allan and Lage-castellanos, Agustin and Moerel, Michelle and Ghosh, Satrajit S and Martino, Federico De},
	doi = {10.7554/eLife.48932},
	journal = {eLife},
	pages = {e48932},
	title = {{Mapping the human subcortical auditory system using histology , postmortem MRI and in vivo MRI at 7T}},
	volume = {8},
	year = {2019}
}

@article{Spratling2017,
	author = {Spratling, M. W.},
	doi = {10.1016/j.bandc.2015.11.003},
	issn = {10902147},
	journal = {Brain and Cognition},
	pages = {92--97},
	publisher = {Elsevier Inc.},
	title = {{A review of predictive coding algorithms}},
	volume = {112},
	year = {2017}
}

@article{Sterzer2018,
	author = {Sterzer, Philipp and Adams, Rick A. and Fletcher, Paul and Frith, Chris and Lawrie, Stephen M. and Muckli, Lars and Petrovic, Predrag and Uhlhaas, Peter and Voss, Martin and Corlett, Philip R.},
	doi = {10.1016/j.biopsych.2018.05.015},
	issn = {18732402},
	journal = {Biological Psychiatry},
	number = {9},
	pages = {634--643},
	pmid = {30007575},
	publisher = {Elsevier Inc},
	title = {{The Predictive Coding Account of Psychosis}},
	volume = {84},
	year = {2018}
}

@article{Tabas2020,
	author = {Tabas, Alejandro and Mihai, Glad and Kiebel, Stefan and Trampel, Robert and {Von Kriegstein}, Katharina},
	doi = {10.7554/ELIFE.64501},
	issn = {2050084X},
	journal = {eLife},
	pages = {1--19},
	pmid = {33289479},
	title = {{Abstract rules drive adaptation in the subcortical sensory pathway}},
	volume = {9},
	year = {2020}
}

@article{Tabas2021,
	author = {Tabas, Alejandro and von Kriegstein, Katharina},
	doi = {10.3389/fncir.2021.644743},
	journal = {Frontiers in Neural Circuits},
	number = {March},
	pages = {1--14},
	title = {{Adjudicating Between Local and Global Architectures of Predictive Processing in the Subcortical Auditory Pathway}},
	volume = {15},
	year = {2021}
}

@article{Tabas2024,
	title = {Fast Frequency Modulation Is Encoded According to the Listener Expectations in the Human Subcortical Auditory Pathway},
	author = {Tabas, Alejandro and Kiebel, Stefan and Marxen, Michael and Von Kriegstein, Katharina},
	year = {2024},
	journal = {Imaging Neuroscience},
	volume = {2},
	pages = {1--20},
	issn = {2837-6056},
	doi = {10.1162/imag_a_00292},
}

@article{Tabas2024a,
	title = {Multiple {{Concurrent Predictions Inform Prediction Error}} in the {{Human Auditory Pathway}}},
	author = {Tabas, Alejandro and Von Kriegstein, Katharina},
	year = {2024},
	journal = {The Journal of Neuroscience},
	volume = {44},
	number = {1},
	pages = {e2219222023},
	issn = {0270-6474, 1529-2401},
	doi = {10.1523/JNEUROSCI.2219-22.2023},
}

@misc{Tabas2022prereg,
	title={Negative prediction error is encoded in the subcotical auditory pathway},
	url={osf.io/86a4x},
	DOI={10.17605/OSF.IO/86A4X},
	publisher={OSF},
	author={Tabas, Alejandro and Sönnichsen, Heike and von Kriegstein, Katharina},
	year={2022},
}

@article{Tustison2010,
	author = {Tustison, N. J. and Avants, B. B. and Cook, P. A. and Zheng, Y. and Egan, A. and Yushkevich, P. A. and Gee, J. C.},
	doi = {10.1109/TMI.2010.2046908},
	issn = {0278-0062},
	journal = {IEEE Transactions on Medical Imaging},
	number = 6,
	pages = {1310-1320},
	shorttitle = {N4ITK},
	title = {N4ITK: Improved N3 Bias Correction},
	volume = 29,
	year = 2010
}

@article{Wacongne2011,
	author = {Wacongne, Catherine and Labyt, Etienne and {Van Wassenhove}, Virginie and Bekinschtein, Tristan and Naccache, Lionel and Dehaene, Stanislas},
	doi = {10.1073/pnas.1117807108},
	issn = {00278424},
	journal = {Proceedings of the National Academy of Sciences of the United States of America},
	number = {51},
	pages = {20754--20759},
	pmid = {22147913},
	title = {{Evidence for a hierarchy of predictions and prediction errors in human cortex}},
	volume = {108},
	year = {2011}
}

@article{Walsh2020,
	author = {Walsh, Kevin S. and McGovern, David P. and Clark, Andy and O'Connell, Redmond G.},
	doi = {10.1111/nyas.14321},
	issn = {17496632},
	journal = {Annals of the New York Academy of Sciences},
	number = {1},
	pages = {242--268},
	pmid = {32147856},
	title = {{Evaluating the neurophysiological evidence for predictive processing as a model of perception}},
	volume = {1464},
	year = {2020}
}

@article{Will2007,
	title = {Brain Wave Synchronization and Entrainment to Periodic Acoustic Stimuli},
	author = {Will, Udo and Berg, Eric},
	year = {2007},
	journal = {Neuroscience Letters},
	volume = {424},
	number = {1},
	pages = {55--60},
	issn = {03043940},
	doi = {10.1016/j.neulet.2007.07.036},
	urldate = {2025-01-12},
	langid = {english},
}

@article{Zhang2001,
	author = {Zhang, Y. and Brady, M. and Smith, S.},
	doi = {10.1109/42.906424},
	issn = {0278-0062},
	journal = {IEEE Transactions on Medical Imaging},
	number = 1,
	pages = {45-57},
	title = {Segmentation of brain {MR} images through a hidden Markov random field model and the expectation-maximization algorithm},
	volume = 20,
	year = 2001
}

  \clearpage
  
  \setcounter{figure}{0}
  \setcounter{table}{0}
  \renewcommand{\thefigure}{S\arabic{figure}}
  \renewcommand{\thetable}{S\arabic{table}}

  \section*{Supplementary materials}
    
\begin{figure}[bth!]
  \centering
  \includegraphics[width=\textwidth]{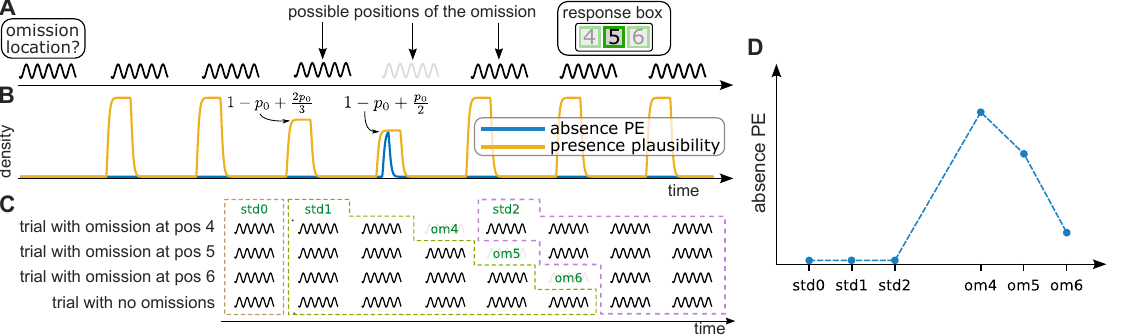}
  \caption{\textbf{Experimental design and hypotheses}. 
  A) Example of a trial. Each trial consisted of a sequence of up to eight repetitions of one pure tone (standards; black); in $p_0 = 6/7$ of the trials one of the pure tones was omitted (omission; light grey). In each trial, the omission could be located in positions 4, 5, or 6 of the sequence. Participants were instructed to report, via a button press, the position of the omission within the sequence. Tones and omissions within each trial were separated by 500\,ms inter-stimulus-intervals ($ISI = 500$\,ms). 
  B) Expected responses from regions encoding absence prediction error (aPE). Predictions are defined as the probability of hearing a tone before the tone is heard. aPE is defined as the predicted probability of the stimulus presence during an omission by an ideal observer.
  C) Definition of the different experimental conditions (i.e., regressors of interest).
  D) Expected responses to each of the regressors defined in (C) for regions encoding absence prediction error (aPE). 
  \label{fig:fulldesign}}
\end{figure}

\begin{table}[bth!]
  \centering
  \begin{tabular}{lrrrrrr}
\toprule
\multicolumn{7}{l}{Cohen's d} \\
                & ICL                  & ICR                  & MGBL                 & MGBR                 & aHGL                 & aHGR                 \\
\midrule
adaptation (std0  $>$ std1)  & \textbf{2.91}        & \textbf{3.6}         & \textbf{1.01}        & \textbf{0.984}       & \textbf{2.4}         & \textbf{2.22}        \\
omissions (om4  $>$ std1)   & 0.427                & \textbf{0.884}       & \textbf{2.29}        & \textbf{1.59}        & \textbf{3.73}        & \textbf{2.49}        \\
\midrule
\\
\multicolumn{7}{l}{$p$-values (corrected)} \\
                & ICL                  & ICR                  & MGBL                 & MGBR                 & aHGL                 & aHGR                 \\
\midrule
adaptation (std0  $>$ std1)  & $6 \times 10^{-5}$   & $4 \times 10^{-5}$   & $0.018$              & $0.024$              & $9 \times 10^{-5}$   & $5 \times 10^{-5}$   \\
omissions (om4  $>$ std1)   & $0.14$               & $0.043$              & $1 \times 10^{-4}$   & $2 \times 10^{-3}$   & $2 \times 10^{-5}$   & $1 \times 10^{-4}$   \\
\midrule
\\
\multicolumn{7}{l}{$p$-values (uncorrected)} \\
                & ICL                  & ICR                  & MGBL                 & MGBR                 & aHGL                 & aHGR                 \\
\midrule
adaptation (std0  $>$ std1)  & $6 \times 10^{-6}$   & $3 \times 10^{-6}$   & $6 \times 10^{-3}$   & $6 \times 10^{-3}$   & $1 \times 10^{-5}$   & $5 \times 10^{-6}$   \\
omissions (om4  $>$ std1)   & $0.14$               & $0.022$              & $2 \times 10^{-5}$   & $3 \times 10^{-4}$   & $2 \times 10^{-6}$   & $1 \times 10^{-5}$   \\
\bottomrule
\end{tabular}

  \caption{\textbf{Statistics for the adaptation ($std0 > std1$) and omission ($om4 > std1$) contrasts}. Bold fonts mark significant effects (as measured by the corrected $p$-values; all effects reported in this table are significant). Uncorrected $p$-values were computed using one-sided ranksum tests. Corrected $p$-values were Holm-Bonferroni corrected for 12 comparisons. IC: inferior colliculus (auditory midbrain); MGB: medial geniculate body (auditory thalamus); PAC: auditory cortex.
  \label{tab:omres}}
\end{table}

\begin{figure}[bth!]
  \centering
  \includegraphics[width=\textwidth]{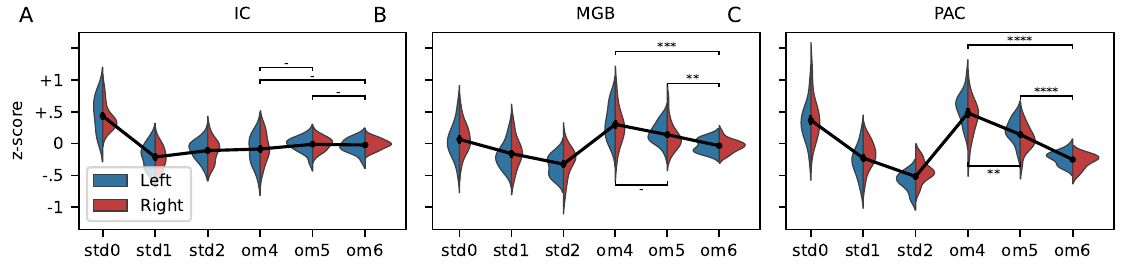}
  \caption{\textbf{Group responses to all experimental conditions}. Estimated BOLD responses to each of the experimental conditions. Responses to the omissions ($om4$, $om5$, $om6$) closely reflect the ideal observer beliefs on the plausibility of the omitted tone. The effect increases qualitatively along the ascending auditory pathway (see below for formal quantitative testing). Violin plots are kernel density estimations across the entire population (one sample per participant). B(L)ue violin plots on the top row show results for the left hemisphere; (R)ed violin plots on the bottom row show the results for the right hemisphere. All $p$-values were computed using one-sided ranksum tests and Holm-Bonferroni corrected for 18 comparisons (nominal corrected and uncorrected $p$-values are listed in Table~\ref{tab:negpe}). 
  $std0$: first standard; $std1$: standards before the omission/unexpected standard; $om4$, $om5$, $om6$: omission in position 4, 5, or 6; $std2$: standards after the omission/unexpected standard (Figure~\ref{fig:fulldesign}C).
  -: not significant; *: $p < 0.05$; **: $p < 0.01$; ***: $p < 0.001$; ****: $p < 0.0001$. IC: inferior colliculus (auditory midbrain); MGB: medial geniculate body (auditory thalamus); PAC: auditory cortex.
  \label{fig:negpe}}
\end{figure}

\begin{table}[bth!]
  \centering
  \begin{tabular}{lrrr}
\toprule
\multicolumn{4}{l}{Cohen's d} \\
                & IC                   & MGB                  & PAC                  \\
\midrule
om4  $>$ om5    & -0.47                & 0.778                & \textbf{1.64}        \\
om4  $>$ om6    & -0.409               & \textbf{1.79}        & \textbf{4.09}        \\
om5  $>$ om6    & 0.122                & \textbf{1.37}        & \textbf{2.8}         \\
\midrule
\\
\multicolumn{4}{l}{$p$-values (corrected)} \\
                & IC                   & MGB                  & PAC                  \\
\midrule
om4  $>$ om5    & $0.84$               & $0.11$               & $2 \times 10^{-3}$   \\
om4  $>$ om6    & $1$                  & $7 \times 10^{-4}$   & $1 \times 10^{-5}$   \\
om5  $>$ om6    & $1$                  & $4 \times 10^{-3}$   & $8 \times 10^{-5}$   \\
\midrule
\\
\multicolumn{4}{l}{$p$-values (uncorrected)} \\
                & IC                   & MGB                  & PAC                  \\
\midrule
om4  $>$ om5    & $0.84$               & $0.027$              & $3 \times 10^{-4}$   \\
om4  $>$ om6    & $0.8$                & $1 \times 10^{-4}$   & $2 \times 10^{-6}$   \\
om5  $>$ om6    & $0.36$               & $8 \times 10^{-4}$   & $1 \times 10^{-5}$   \\
\bottomrule
\end{tabular}

  \caption{\textbf{Statistics for the aPE contrast}. We quantified the strength of the aPE responses as the contrast between the responses to $om4$ and $om6$. The effect increases qualitatively along the ascending auditory pathway (see below for formal quantitative testing).
  Bold fonts mark significant effects (as measured by the corrected $p$-values). Uncorrected $p$-values were computed using one-sided ranksum tests with 15 samples (one sample per participant). Corrected $p$-values were Holm-Bonferroni corrected for 24 comparisons. 
  $std0$: first standard; $std1$: standards before the omission/unexpected standard; $om4$, $om5$, $om6$: omission in position 4, 5, or 6; $std2$: standards after the omission/unexpected standard (Figure~\ref{fig:fulldesign}C).
  IC: inferior colliculus (auditory midbrain); MGB: medial geniculate body (auditory thalamus); PAC: auditory cortex. aPE: absence prediction error.
  \label{tab:negpe}}
\end{table}

\begin{figure}[p]
  \centering
  \includegraphics[width=\textwidth]{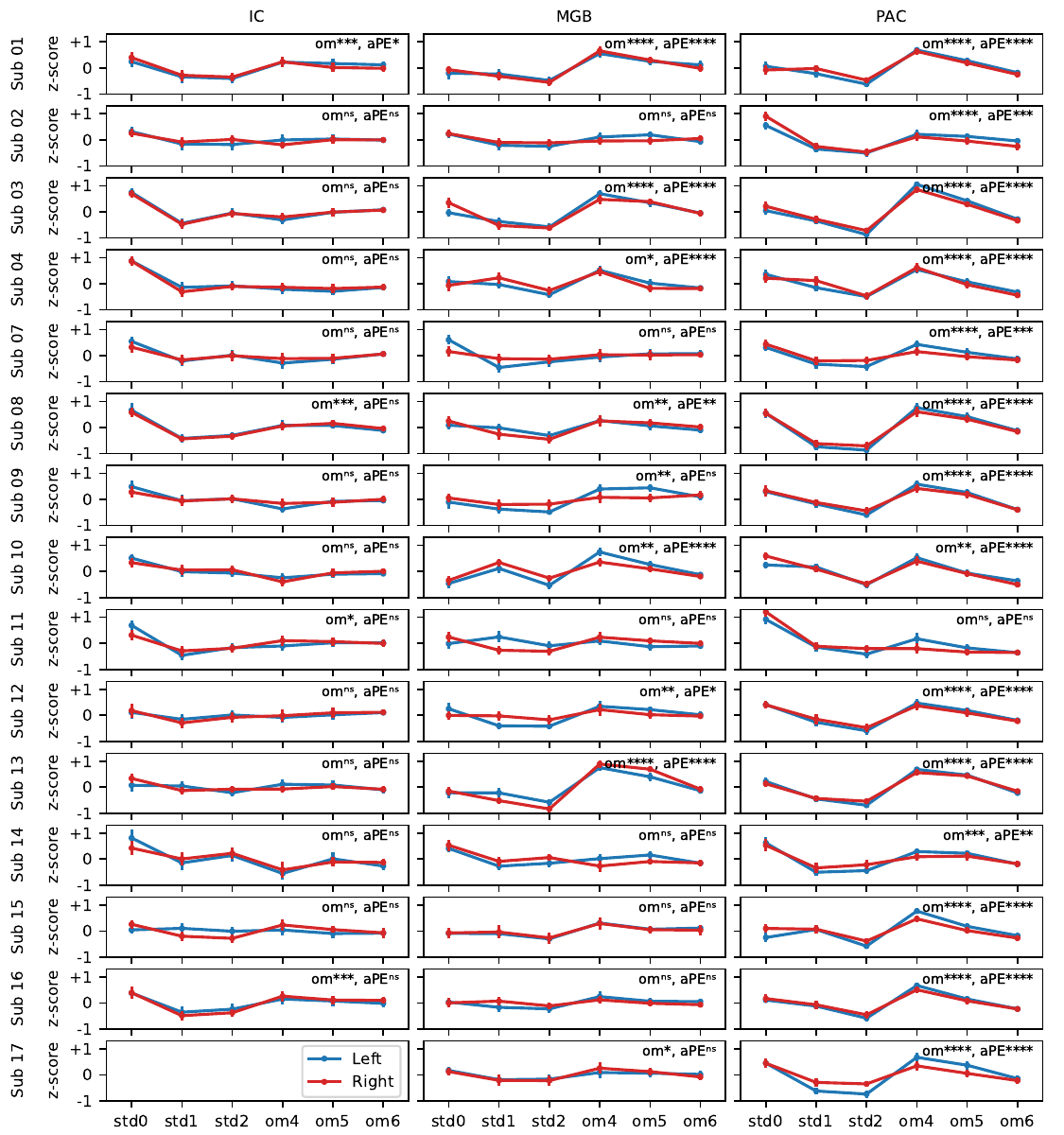}
  \caption{\textbf{Single Participant Results}. Average BOLD responses to each of the experimental conditions for each participant and ROI. Group-level results are qualitatively replicated in most of the participants (cf. Supplementary Figure~S2). B(L)ue lines represent results for the left hemisphere; (R)ed lines represent the results for the right hemisphere. Significance was tested independently for each participant of the omissions contrast (\emph{om}: $om4 > std1$) and the strongest of the aPE contrasts (\emph{pe}: $om4 > om6$) using paired one-sided ranksum tests (each sample was the average response in one run). Data for the auditory midbrain of the 17th participant is missing because the MRI-slab did not cover the midbrain of the participant. All $p$-values were corrected for 6 comparisons independently for each participant. ns: not significant; *: $p < 0.05$; **: $p < 0.01$; ***: $p < 0.001$; ****: $p < 0.0001$. IC: inferior colliculus (auditory midbrain); MGB: medial geniculate body (auditory thalamus); PAC: auditory cortex. aPE: absence prediction error.
  \label{fig:participants}}
\end{figure}

\begin{figure}[bth!]
  \centering
  \includegraphics[width=\textwidth]{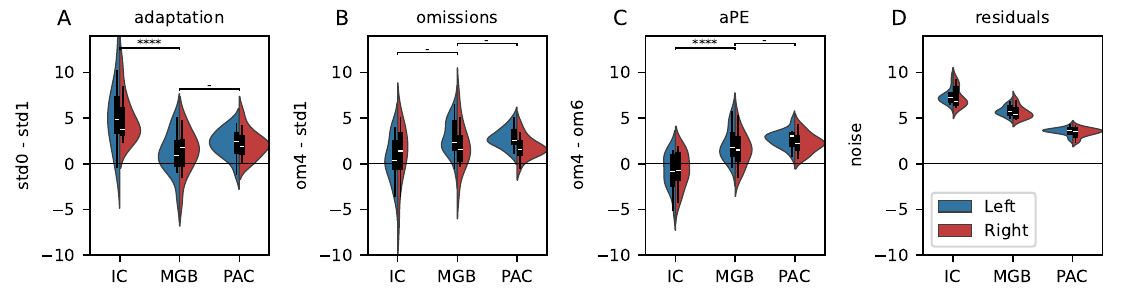}
  \caption{\textbf{Adaptation, omission, and aPE responses normalised by the inverse of the noise of the residuals}. A) Noise of the residuals as measured by the standard deviation of the residuals of the functional localiser after regressing out task-related and nuisance regressors. B-D) Adaptation (B, $std0 > std1$), omissions (C, $om4 > std1$) and aPE (D, $om4 > om6$) contrasts, voxel-wise normalised by the inverse of the residuals noise, and averaged across all voxels. Violin plots are kernel density estimations across the entire population (one sample per participant). B(L)ue halves of the violin plots correspond to the left hemisphere; (R)ight halves to the right hemisphere. All $p$-values were computed using one-sided ranksum tests and Holm-Bonferroni corrected for 6 comparisons (nominal corrected and uncorrected $p$-values are listed in Table~\ref{tab:emergence}). *: $p < 0.05$; **: $p < 0.01$; ***: $p < 0.001$; ****: $p < 0.0001$. IC: inferior colliculus (auditory midbrain); MGB: medial geniculate body (auditory thalamus); PAC: auditory cortex. aPE: absence prediction error.
  \label{fig:residuals}}
\end{figure}

\begin{table}[bth!]
  \centering
  \begin{tabular}{lrrr}
\toprule
\multicolumn{4}{l}{Cohen's d} \\
                & adaptation (std0-std1) & omissions (om4-std1)   & absence PE (om4-om6)          \\
\midrule
IC $\neq$ MGB   & \textbf{1.04}        & \textbf{-0.782}      & \textbf{-1.43}       \\
MGB $\neq$ aHG  & 0.21                 & \textbf{0.813}       & 0.499                \\
\midrule
\\
\multicolumn{4}{l}{$p$-values (corrected)} \\
                & adaptation (std0-std1) & omissions (om4-std1)   & absence PE (om4-om6)          \\
\midrule
IC $\neq$ MGB   & $4 \times 10^{-4}$   & $0.042$              & $4 \times 10^{-6}$   \\
MGB $\neq$ aHG  & $0.54$               & $0.021$              & $0.24$               \\
\midrule
\\
\multicolumn{4}{l}{$p$-values (uncorrected)} \\
                & adaptation (std0-std1) & omissions (om4-std1)   & absence PE (om4-om6)          \\
\midrule
IC $\neq$ MGB   & $7 \times 10^{-5}$   & $0.014$              & $6 \times 10^{-7}$   \\
MGB $\neq$ aHG  & $0.54$               & $5 \times 10^{-3}$   & $0.12$               \\
\bottomrule
\end{tabular}

  \caption{\textbf{Statistics for the difference of effect strength between regions}. Bold fonts mark significant effects (as measured by the corrected $p$-values). Uncorrected $p$-values were computed using two-sided ranksum tests. Corrected $p$-values were Holm-Bonferroni corrected for 6 comparisons. IC: inferior colliculus (auditory midbrain); MGB: medial geniculate body (auditory thalamus); PAC: auditory cortex. aPE: absence prediction error.
  \label{tab:emergence}}
\end{table}

\end{document}